\documentclass[a4paper,fleqn,usenatbib]{mnras}

\usepackage{newtxtext,newtxmath}

\usepackage[T1]{fontenc}
\usepackage{ae,aecompl}

\usepackage[dvipdfmx]{graphicx}	
\usepackage{amsmath}	
\usepackage{amssymb}	
\usepackage{indentfirst}	
\usepackage{pict2e}
\usepackage{color}
\usepackage{bm}
\usepackage{tablefootnote}
\usepackage{pict2e}
\usepackage{color}
\usepackage{ulem}

\title[Circular polarization of near infrared light]{
High circular polarization of near infrared light induced by micron-size dust grains
}

\author[Fukushima et al.]{
Hajime Fukushima$^{1}$\thanks{E-mail:fukushima@ccs.tsukuba.ac.jp},
Hidenobu Yajima$^{1}$,
Masayuki Umemura$^{1}$
\\
$^{1}$Center for Computational Sciences, University of Tsukuba, Ten-nodai, 1-1-1 Tsukuba, Ibaraki 305-8577, Japan\\
}
\date{Accepted XXX. Received YYY; in original form ZZZ}

\pubyear{2019}

\begin{document}
\label{firstpage}
\pagerange{\pageref{firstpage}--\pageref{lastpage}}
\maketitle

\begin{abstract}
We explore the induction of circular polarization (CP) of near-infrared light in star-forming regions using three-dimensional radiative transfer calculations. The simulations trace the change of Stokes parameters at each scattering/absorption process in a dusty gas slab composed of aligned grains. 
We find that the CP degree enlarges significantly according as the size of dust grains increases
and exceeds $\sim 20$ percent for micron-size grains. 
Therefore, if micron-size grains are dominant in a dusty gas slab,
the high CP observed around luminous young stellar objects can be accounted for.  
The distributions of CP show the asymmetric quadrupole patters regardless of the grain sizes. 
Also, we find that the CP degree depends on the relative position of a dusty gas slab. 
If a dusty gas slab is located behind a star-forming region,  
the CP reaches $\sim 60$ percent in the case of $1.0~{\rm \mu m}$ dust grains.
Hence, we suggest that the observed variety of CP maps can be explained by different size distributions of dust grains and the configuration of aligned grains. 
\end{abstract}

\begin{keywords}

\end{keywords}


\section{Introduction}\label{introduction}

Recent observations show that circular polarization (CP)  of near infrared-light exists around young stellar objects \citep[YSOs; e.g.,][]{2010OLEB...40..335F,2013ApJ...765L...6K,2014ApJ...795L..16K, 2016AJ....152...67K, 2018AJ....156....1K}. In many cases, the observed CP maps exhibit a quadrupole pattern. Although the degree of CP is at a level of $\lesssim 3$ percent in most of the cases, a few YSOs show CP higher than 10 percent.  In particular,  the star-forming region NGC6334-V exhibits $\sim 20$ percent CP \citep{2013ApJ...765L...6K}. However, the origin of such high CP has not revealed.  

It is argued that interstellar circularly polarized light could be responsible for 
the enantiomeric excesses of biological molecules \citep{1998Sci...281..672B}.
The homochirality of biological molecules on the Earth is a long-standing mystery.
In modern astrobiology, seeking for the origin of  homochirality in space is a significant theme. 
It is reported that various amino acids exist in meteorites, and also, the enantiomeric excesses of L-amino acids are detected in the Murchison and Mukundpura meteorites \citep{1997Natur.389..265E,2018P&SS..164..127P}. It has been hitherto proposed that amino acids were formed in interstellar space and delivered onto the Earth by meteorites \citep{1991OLEB...21...59B}. 
As suggested in \citet{1998Sci...281..672B}, CP of interstellar radiation could break down the symmetry of chiral amino acids efficiently, resulting in the enantiomeric excess. 

Previous works have shown that CP can be produced by the scattering from non-spherical dust grains \citep{2000MNRAS.314..123G,2002A&A...385..365W}. 
\citet{2005OLEB...35...29L} argued that the high degree of CP than $\sim 20$ percent cannot be produced by single-scattering. 
Therefore,  the dichroic extinction has been frequently considered as the origin of CP \citep[e.g.,][]{2013ApJ...765L...6K}. 
However, their analyses have been made for limited physical conditions of dust grains.
There is a wide range of physical conditions, e.g., different grain sizes, dust distributions around a YSO, and so on. 
In particular, dust grains larger than $0.75~{\rm \mu m}$ have not been considered in the previous works.

Recent observations suggest that dust grains grow through accretion and coagulation in high-density regions near massive stars and reach micron size \citep{2010Sci...329.1622P, 2010A&A...511A...9S}. 
If the dust size is close to the wavelengths of photons, the change of polarization on scattering becomes significantly different from that in cases of smaller grains.  
However, CP in the environment with such large dust grains has not been investigated.  
Therefore, in this {\it Letter}, we study the generation of CP for a wide range of grain sizes by three-dimensional radiative transfer simulations. Especially, we concentrate on the impacts of grain sizes 
on the degree and pattern of CP. Also, we explore the dependence of CP 
on the spatial distributions of aligned grains around a star-forming region.  

\section{Numerical method}\label{method}
\begin{figure}
\begin{center}
\includegraphics[width=\columnwidth]{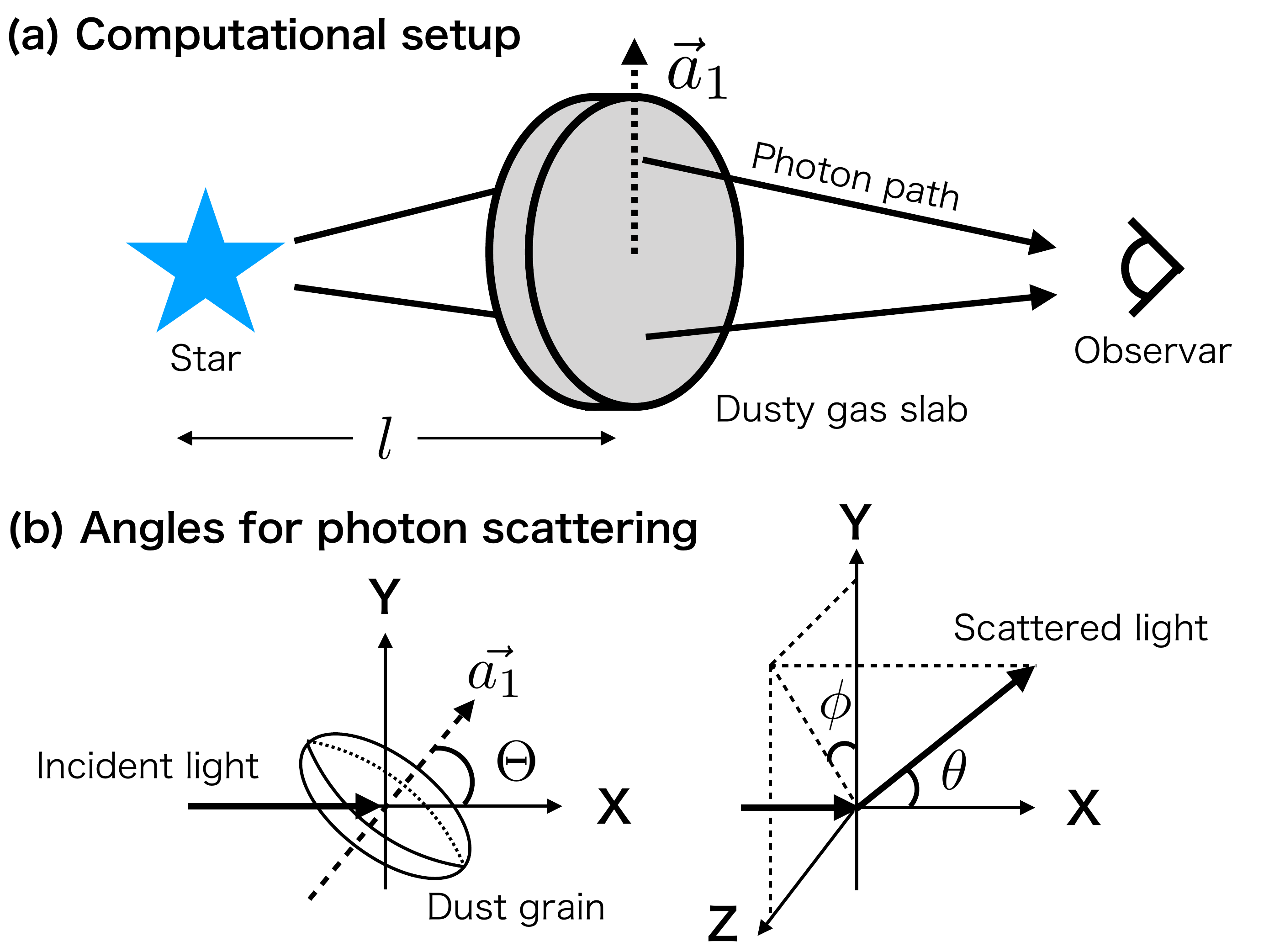}
\end{center}
\caption{
Schematic view of our radiative transfer calculation: 
(a) the spatial positions of the radiation source, the dusty gas slab, and the observer; and (b) definitions of angles related to single scattering.  
The vector $\vec{a}_{1}$ and $l$ represent respectively the direction of the dust minor axis and the distance between the star and the dusty gas slab.
In the fiducial model, we set the star in the center of the slab as $l=0$.
In the panel (b), $\Theta$ and $(\theta, \phi)$ represent the opening angle between the incident light and the minor axis of the dust grains and the scattered direction \citep[see,][]{2013arXiv1305.6497D}.
}
\label{schematic_picture_of_simulation}
\end{figure}

We develop a numerical code for calculating radiation transfer of the Stokes parameters based on the Monte Carlo technique. 
Here, we consider the near-infrared wavelength of $\lambda=2.14~{\rm \mu m}$, in which the high degree of CP is detected in observations \citep{2013ApJ...765L...6K}.
We set a circular slab of dusty gas with a radius of $1~{\rm pc}$ and a thickness of $0.2~{\rm pc}$ near a massive star.
In this {\it Letter}, we study the generation mechanism of high CP degree with this idealized setup.
Note that, in reality, the spatial distribution of dusty gas around a star can be complicated with clumps, holes, and shell-like structure.

In a fiducial model, the star is located in the center of the slab.
To study the impact of the relative position between the star and the dusty slab, we consider the cases in which the dusty slab is placed at  $0.1~{\rm pc}$ behind or in front of the star (Fig. \ref{schematic_picture_of_simulation}-a). 
The number density of dust grains is set as the scattering optical depth along the horizontal direction is unity at $\lambda=2.14~{\rm \mu m}$.
If the optical depth is larger than unity, the number of multiple scattering photons increases. 
In this case, due to mismatch of the sign of CP, the CP degree is likely to be decreased. 
Here, photon packets are emitted from the star and assumed to be unpolarized initially.
For the radiative transfer calculations, the dusty slab is divided by $300 \times 300 \times 60$ cells. 

In the transport of a photon packet, the path length from a scattering position to next scattering one is evaluated by the optical depth that is stochastically determined by $\tau = - \ln (R)$, where $R$ is the random number in the range of $0 - 1$. 
We pursue $10^{7}$ photon packets, which can satisfactorily reduce the shot noise from the random number. 

Each photon packet has information of the Stokes vector $\bm{I} = (I, Q, U, V)^{\mathrm T}$.
The transformation of the Stokes vector in one scattering event is determined by the M\"{u}ller matrix as \citep[e.g.,][]{1983asls.book.....B}, 
\begin{eqnarray}
\left( \begin{array}{c} 
I\\
Q \\ 
U \\
V
\end{array} \right)_{\rm sc} =  \frac{1}{k^2r^2} \left(
\begin{matrix} 
S_{11}  &  S_{\rm 12} & S_{\rm 13} & S_{\rm 14} \\
S_{21}  &  S_{\rm 22}& S_{23} & S_{24} \\ 
S_{31}  &  S_{\rm 32}& S_{33} & S_{34} \\ 
S_{41}  &  S_{\rm 42}& S_{43} & S_{44}
\end{matrix}
\right)
\left( \begin{array}{c} 
I \\
Q \\ 
U \\
V
\end{array} \right)_{\rm i}, \label{1.1}
\end{eqnarray}
where $k$ and $r$ are the photon wavenumber and the distance from the scattering point.
Using Equation \eqref{1.1}, the differential cross section for scattering is estimated as
\begin{eqnarray}
\frac{dC_{\rm sca}}{d \Omega} = \frac{1}{k^2} \frac{ ( S_{11} I + S_{12} Q + S_{13} U + S_{14} V) }{I}. \label{1.2}
\end{eqnarray}
We obtain the cross section by integrating Equation \eqref{1.2} with respect to solid angle $\Omega$.
The direction of scattering light is stochastically chosen based on the phase function obtained from Equation \eqref{1.2}.
Note that the Stokes parameters ($Q$ and $U$) depend on the coordinate. 
In Equation \eqref{1.1}, the direction of the linear polarization of $(Q, U) = (1, 0)$ corresponds to the plane which contains incident and scattering directions.
We also include the absorption of photons estimated as \citep[e.g., manual for RADMC-3D,][]{2012ascl.soft02015D},
\begin{eqnarray}
\frac{d}{ds} \left( \begin{array}{c} 
I_{\rm h}\\
I_{\rm v} \\ 
U \\
V
\end{array} \right) =  - \left(
\begin{matrix} 
\alpha_{\rm h}  &  0 & 0 & 0 \\
0 &  \alpha_{\rm v}& 0 & 0 \\ 
0 &  0 & \alpha_{\rm uv} & 0 \\ 
0 &  0 & 0 & \alpha_{\rm uv}
\end{matrix}
\right)
\left( \begin{array}{c} 
I_{\rm h} \\
I_{\rm v} \\ 
U \\
V
\end{array} \right), \label{1.1.12}
\end{eqnarray}
where $I_{\rm h} = (I+Q)/2$ and $I_{\rm v} = (I-Q)/2$.
In Equation \eqref{1.1.12}, $\alpha_{\rm h}$ and $\alpha_{\rm v}$ are absorption coefficients for the horizontal and vertical polarization light for the scattering plane, and $\alpha_{\rm uv} = (\alpha_{\rm h}+\alpha_{\rm v})/2$.

We evaluate the M\"{u}ller matrix using the DDSCAT code \citep{1994JOSAA..11.1491D}, where dust grains are represented as an array of electric dipoles.
Given that a dust grain is much smaller than the wavelength of incident light ($2\pi a_{\rm d} \ll \lambda$),
the electromagnetic field in the dust grain can be understood as  
the behavior of a couple of dipoles along intersecting axes at right angles on a plane perpendicular to the incident light, i.e., Rayleigh approximation. 
On the other hand, according as the grain size increases, the interaction among multiple dipoles leads to the complicated phase distributions of circular polarization. 
In this case, the polarization of scattered light is given by superposition of electromagnetic waves from all dipoles. 
We discrete the opening angle $\Theta$ with 91 bin, and the scattered angles $(\theta, \phi)$ with 1024 bin (see Fig. \ref{schematic_picture_of_simulation}) by HEALPix \citep{2005ApJ...622..759G}.

In this study, we assume dust grains of oblate spheroids with the axial ratio of 2:1.
The oblate dust grains are preferred rather than prolate ones, and this axis ratio is similar to that was suggested for the aligned interstellar dust grains \citep{1995ApJ...450..663H, 2002ApJ...574..205W}. 
The dust grains are placed as the minor axes of grains are aligned with the radial direction of the dusty slab. 
In diffuse interstellar medium, dust grains larger than $a_{\rm d} \sim 0.04 \mu {\rm m}$ are aligned \citep{1995ApJ...444..293K, 2015ARA&A..53..501A}.
The CP degree is mainly generated by the larger grains ($a_{\rm d} \geqq 0.1~{\rm \mu m}$) in this study, and thus we adopt this simple assumption. 
We use the dielectric function of astronomical silicate \citep[$m= 1.7+0.034 i$, ][]{2003ApJ...598.1026D} to calculate the M\"{u}ller matrix.

The size distributions $n_{\rm d}$ of dust grains in the diffuse interstellar medium is often modeled as the MRN mixture ($n_{\rm d} \propto a_{\rm d}^{-3.5}$) in the range $0.005 {\rm \mu m} \leqq a_{\rm d} \leqq 0.25
 {\rm \mu m}$, where $a_{\rm d}$ is the grain size  \citep{1977ApJ...217..425M}.
Around YSOs, on the other hand, the size distributions can differ from the MRN model because the grain size increases via metal accretion and coagulation \citep[e.g.,][]{2009A&A...502..845O, 2013MNRAS.434L..70H}.
The mid-infrared scattered light from dense cores in star-forming regions suggests that there are micron-sized dust grains \citep{2010Sci...329.1622P, 2010A&A...511A...9S}. 
However, the grain size distributions around YSOs are still unclear. 
Here, we employ three models with a single size of $0.1$, $0.3$ or $1.0 ~{\rm \mu m}$ 
as well as the MRN mixture case in the size range of $10^{-3} {\rm \mu m} \leqq a_{\rm d} \leqq 1 {\rm \mu m}$.
 In this work, the grain size is defined as $a_{\rm d} = \left( 3V/ 4 \pi \right)^{1/3}$, where $V$ is the volume of the oblate dust.
The dust grains are assumed to be composed of dielectric materials.
We note that the albedos for $a_{\rm d} = 0.1$, $0.3$ and $1.0~{\rm \mu m}$ are $0.1$, $0.7$ and $0.9$.

Finally, we construct the observational images on the screens in arbitrary directions
using the Stokes vectors at scattering points.
For the purpose, we solve the radiative transfer to the observer as \citep[e.g.,][]{2012MNRAS.424..884Y}
\begin{eqnarray}
 	\bm{I}_{\rm ob} = \bm{I}_{\rm sc} \exp{\left[-(\tau_{\rm ab}+\tau_{\rm sc}) \right]}, \label{0123.1}
\end{eqnarray}	
where $\bm{I}_{\rm ob}$ and $\bm{I}_{\rm sc}$ are the Stokes vectors at the observer and at a scattering point, $\tau_{\rm ab}$ and $\tau_{\rm sc}$ are the optical depths of absorption and scattering from the scattering point to the observer.
By integrating the Stokes vectors calculated in Equation \eqref{0123.1}, the polarization map is built up. 
In this work,  the propagation angle of photon packets to the external screen is discretized by HEALPix \citep{2005ApJ...622..759G} with 1024 angle bins.

\section{Results}
\begin{figure*}
\begin{center}
\includegraphics[width=160mm]{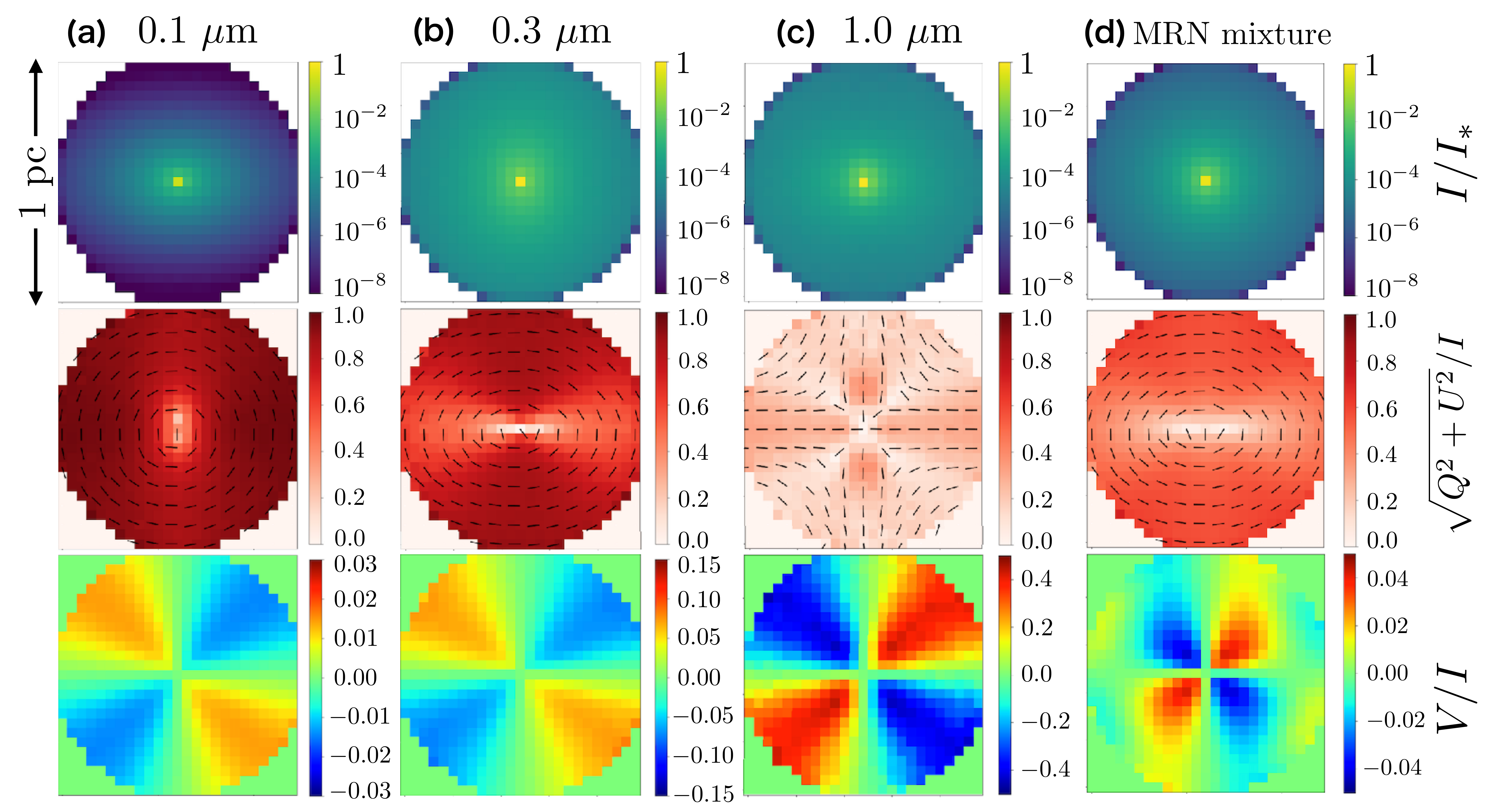}
\end{center}
\caption{
The distributions of intensity (top row), linear polarization (middle) and circular polarization (bottom) in the face-on view to the dusty slab.
The four columns of panels represent the results for (a) $0.1~{\rm \mu m}$, (b) $0.3~{\rm \mu m}$, (c) $1.0~{\rm \mu m}$, and (d) the MRN mixture.
In the top panels, we plot the intensity ratio of the scattered light to the stellar direct light ($I_{*}$). 
In the middle panels, the directions of polarization are also shown by bars.}
\label{zu1}
\end{figure*}

\begin{figure}
\begin{center}
\includegraphics[width=\columnwidth]{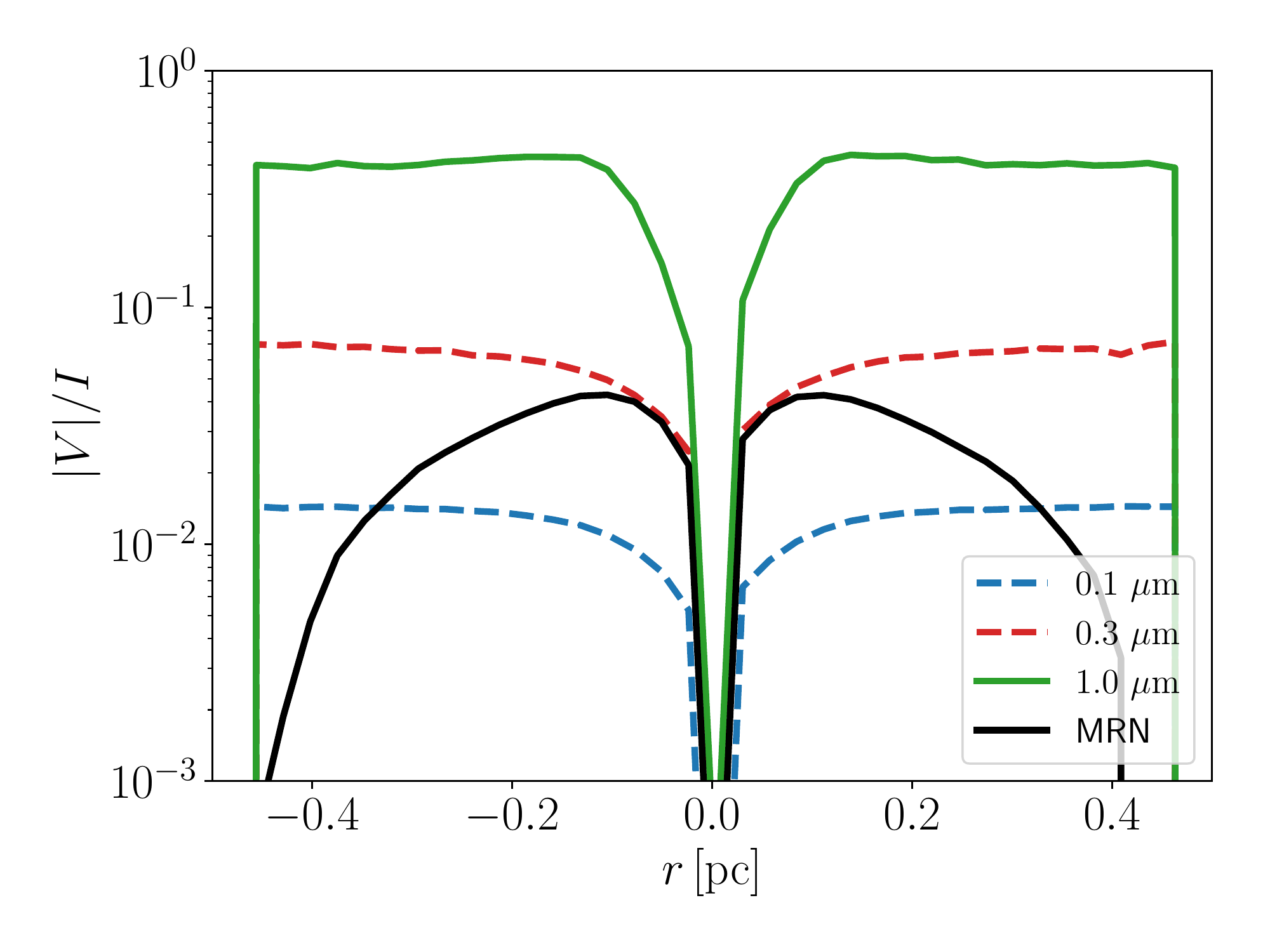}
\end{center}
\caption{ 
The absolute value of CP degree along the $45^{\circ}$ direction in the observation maps.
Blue, red, green and black lines correspond to the cases at $a_{\rm d} = 0.1$, $0.3$, $1.0~{\rm \mu m}$ and the MRN mixture.
The solid and dashed lines represent $V/I>0$ or $V/I< 0$ cases, respectively. 
}
\label{zu3}
\end{figure}
 \begin{figure}
 \begin{center}
 \includegraphics[width=\columnwidth]{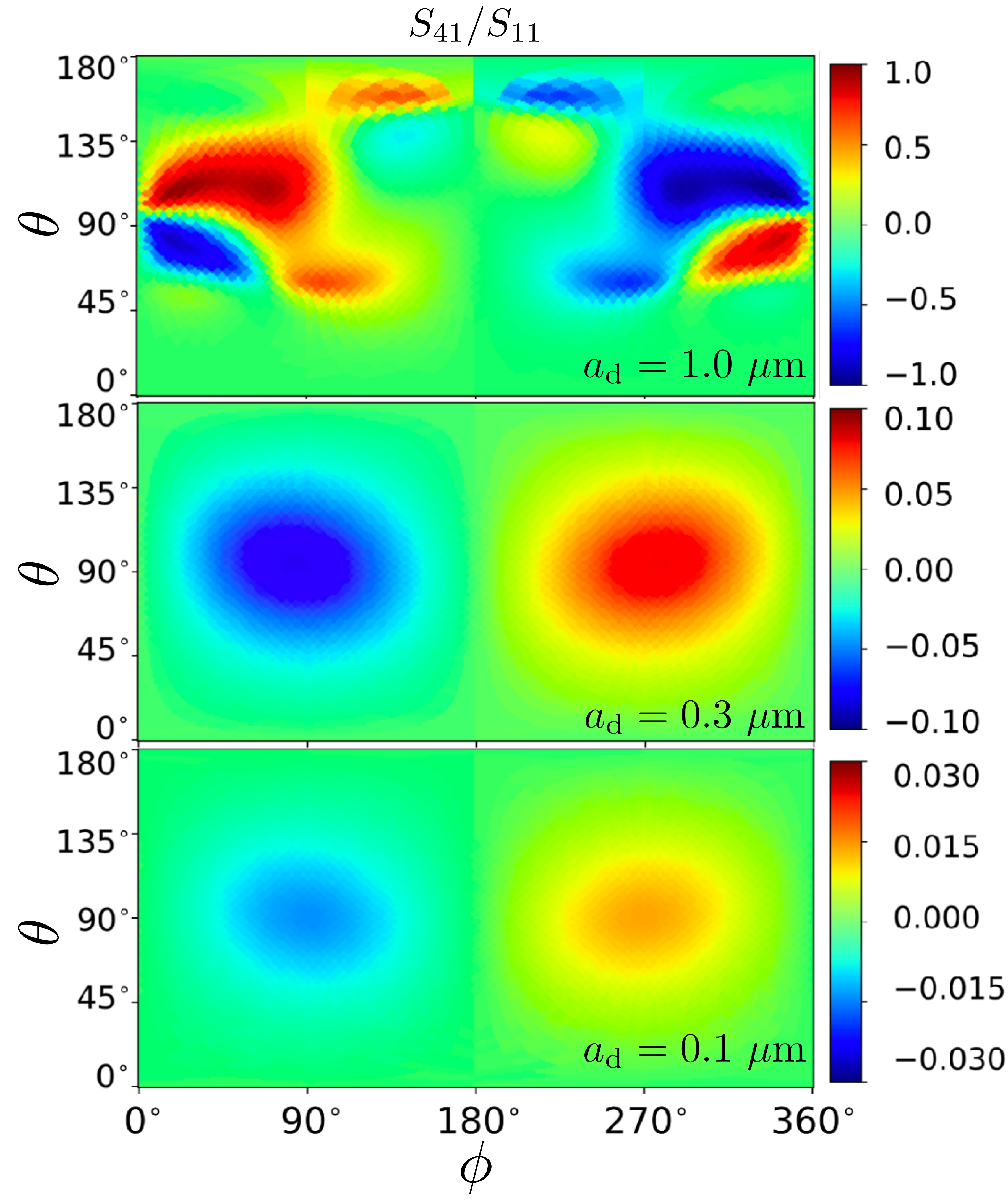}
 \end{center}
 \caption{ 
 The distributions of $S_{41}/S_{11}$ in the scattering plane of $(\phi, \theta)$ 
 for the cases with $a_{\rm d} = 1.0$, $0.3$ and $0.1~{\rm \mu m}$.
  }
 \label{zu2}
 \end{figure}
 \begin{figure}
 \begin{center}
 \includegraphics[width=\columnwidth]{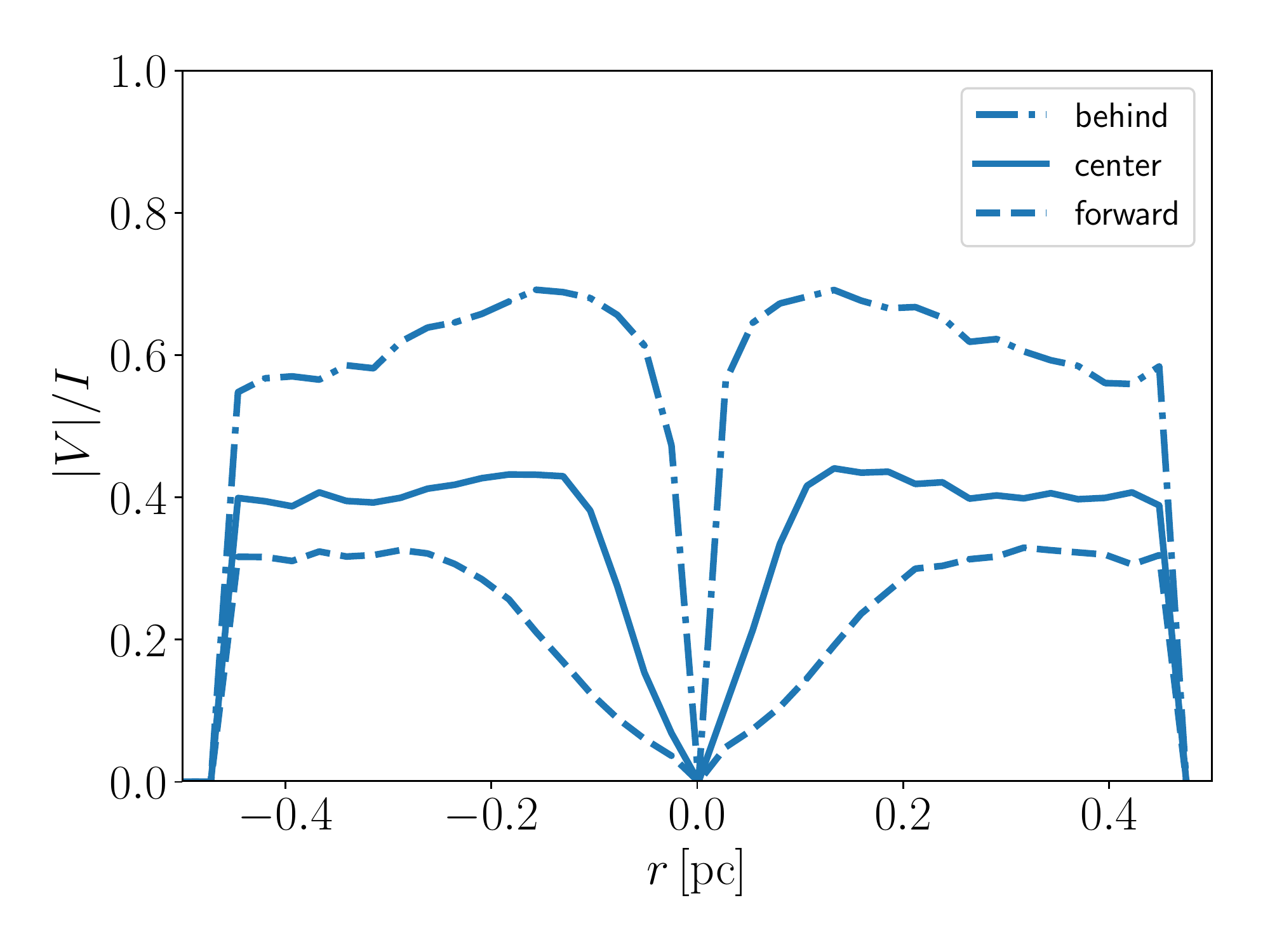}
 \end{center}
 \caption{ 
 Same as Figure \ref{zu3}, but for the cases where the slab center is at $l=0.1$ (dot), $0.0$ (solid) and $-0.1~{\rm pc}$ (dot-dashed, respectively) at $a_{\rm d}=1.0~{\rm \mu m}$.
  }
 \label{zu4}
 \end{figure} 

The resultant CP maps for different grain size models are shown in Figure \ref{zu1}. In this work, we present the face-on views of the dusty gas slab. 
\footnote{The details of the dependence on viewing angles and dust properties will be presented in our forthcoming paper (Fukushima et al. in prep.).} 
The top row shows the ratio of the emergent intensity $I$ to the intrinsic intensity $I_*$ emitted from the star, the middle does linear polarization, and the bottom does  circular polarization.
According as the grain size becomes small, the scattering albedo decreases, and the absorption optical depth increases. Therefore, in the case of $a_{\rm d} = 0.1~{\rm \mu m}$,  the emergent intensity  is as low as $\sim 10^{-8} I_*$ at the edge of the slab. 
For $a_{\rm d} \geqq 0.3 ~{\rm \mu m}$, the albedo is close to unity, and diffuse photons can propagate even at the edge. 
In the case of the MRN model, small grains with low scattering albedos absorb photons efficiently, resulting in the suppression of photon propagation over a long distance. 

The distributions and directions of linear polarization are shown in the middle row in Figure \ref{zu1}.
In the cases with $a_{\rm d} = 0.1$ and $0.3~{\rm \mu m}$, the maximum linear polarization degree reaches  $\sim 80$ percent
and the polarization vectors exhibit the concentric pattern.
In both cases, 
the photon scattering is understood by the Rayleigh approximation  ($2 \pi a_{\rm d} \ll \lambda$),
in which the scattered light is linearly polarized along the vertical plane to incident and outgoing vectors. 
On the other hand, as the grain size is close to $a_{\rm d} = 1.0~{\rm \mu m}$, the linear polarization pattern becomes different from the Rayleigh limit. 
If photons propagate along the face-on or edge-on direction to $1.0~{\rm \mu m}$ grains and they are scattered to the observer, the oscillation of electric dipole moment along the parallel plane to the incident and outgoing vectors becomes dominant. Thus, the polarization vectors point to  the radial directions. 
If the grain size is much larger than the wavelength of photons, the direction of linear polarization is aligned with the major axis of the oblate dust grain regardless of the incident angle.    

The bottom row of Figure \ref{zu1} shows the CP maps, i.e., $|V|/I$.
The angular distributions show the symmetric quadrupole patterns.
These patterns nicely match recent observations \citep[e.g.,][]{2014ApJ...795L..16K}.
Compared to the small dust grains ($0.1$ and $0.3 ~{\rm \mu m}$), the sign of CP is reversed in the cases of $a_{\rm d} = 1.0~{\rm \mu m}$ and the MRN mixture.
In all cases, the absolute value of CP becomes maximum in the diagonal direction. 
In Figure \ref{zu3}, we show the degrees of CP along $45^{\circ}$ direction of the slab as a function of distance from the star. Here, $0^{\circ}$ is defined as the direction of the minor axis of dust grains. 
The CP degree increases as the grain size becomes larger, and it reaches $40$ percent for $a_{\rm d} = 1.0~{\rm \mu m}$.
 The CPs for the models of single size grains are roughly constant.  
On the other hand, in the case of the MRN mixture, $|V|/I$ decreases as the radial distance increases.
Because of the contribution of small dust grains, the absorption optical depth is larger than the single size models. 
After the absorption of photons, penetrating photons are linearly polarized. The direction of the linear polarization is 
perpendicular to that of photons after scattering. 
According to the M\"{u}ller matrix, the sign of $S_{42}$ is inverse to $S_{43}$. 
Therefore, if photon packets after absorption and scattering do the final interaction with dust, and they are collected in the same pixel on the imaging screen, 
the CP becomes weak due to the offset. 
This is why the CP decreases at the large radius in the case of the MRN. 

In the MRN mixture, the CP degree cannot become higher than $5$ percent, although micron-sized grains are contained.
As shown in the single size models, the CP from the scattering on large dust grains with $a_{\rm d} \gtrsim 0.3~{\rm \mu m}$ offsets that by small grains, resulting in the low level of CP. 
If the typical dust size is as large as $\sim 1 \rm \mu m$ through accretion and coagulation, the CP degree becomes higher as a result of reduced offset. 
Recent observations indicate that the CPs in star-forming regions are mostly
at a level of less than $\sim 5$ percent, 
while some observations show the CP degree as high as $\sim 20$ percent \citep{2014ApJ...795L..16K}. 
We suggest that the diversity of observed CP maps can originate from different size distributions
of dust grains. 
Actually, the observations of $3.6~{\rm \mu m}$ emission from dense cores in the star-forming clouds  support $1~{\rm \mu m}$ dust grains \citep{2010A&A...511A...9S, 2010Sci...329.1622P}, and metal accretion or coagulation is thought to induce such grain growth \citep[e.g.,][]{2009A&A...502..845O,2013MNRAS.434L..70H}. 
Also, the positive correlation of the CP degree with the luminosity of YSOs is shown in \citet{2014ApJ...795L..16K}.
The high-density regions are preferred for the formation sites of massive stars \citep[e.g.,][]{2003ApJ...585..850M,2018MNRAS.473.4754F}, and such environments may enhance the dust growth.
Note that, however, the growth rate and size distributions of dust grains are still under debate, 
because they are sensitive to physical conditions in molecular clouds that are not understood well. 

The distributions and degrees of CP are closely linked with the M\"{u}ller matrix.
In particular, $S_{41}$ component produces the CP from non-polarized light, and mainly determines the pattern of CP as shown in Figure \ref{zu1} \citep[e.g.,][]{2000MNRAS.314..123G}.
In Figure \ref{zu2}, we show the distributions of $S_{41}/S_{11}$ against scattering angles for the cases that the opening is $\Theta = 45^{\circ}$. 
In this study, we assume oblate dust grains, and thus the phase function of each component of the M\"{u}ller matrix becomes antisymmetric to $\phi$ with respect to the center $\phi = 180^{\circ}$.
In the case of $a_{\rm d} = 0.1 ~\rm \mu m$, the $S_{41}$ component is estimated based on the Rayleigh approximation as 
 \citep[e.g.,][]{2000MNRAS.314..123G, 2017ApJ...839...56T}
\begin{eqnarray}
	S_{41} \propto \left( \alpha_1 \alpha_3^{*} - \alpha_{1}^{*} \alpha_3 \right) \sin \theta \sin \phi \sin 2 \Theta. \label{1.3}
\end{eqnarray}
Therefore, $S_{41}/S_{11}$ is symmetric to $\theta$ with respect to the center $\theta = 90^{\circ}$.
In this case, the CP becomes maximum if the scattering angle is $(\theta, \phi) = (90^{\circ}, 90^{\circ})$. 
In addition, $S_{41}$ becomes maximum for $\Theta = 45^{\circ}$ or $135^{\circ}$. This reflects the angular distributions of CP as in Figure \ref{zu1}.

As the grain size increases, the distributions of $S_{41}$ become different from equation (\ref{1.3}). 
With $a_{\rm d} = 1.0~{\rm \mu m}$, the peak value of $S_{41}/S_{11}$ is around unity, and much larger than that for $0.1 ~\rm \mu m$ grains, and the distributions in the phase space are obviously different.
In particular, the sign is inverse around $(\theta, \phi) = (90^{\circ}, 90^{\circ})$.
These differences result in the different CP distributions between the cases for small and large dust grains as shown in Figure \ref{zu1}.
For the MRN model, the pattern of the sign is similar to that for $1.0~{\rm \mu m}$ dust grains.
In the Rayleigh approximation, the scattering coefficient is proportional to the grain size with the dependence of $a_{\rm d}^4$. The abundance of dust grains is given as $\propto a_{\rm d}^{-3.5}$ in the MRN mixture. 
By multiplying the scattering coefficient of the geometrical cross section ($\pi a_{\rm d}^2$) and the MRN size distribution ($\propto a_{\rm d}^{-3.5}$), the scattering optical depth per a specific grain size bin is estimated as $d\tau/da_{\rm d} \propto a_{\rm d}^{2.5}$.
Therefore, in the MRN model, the contribution of the larger dust grains ($\sim 1~\rm \mu m$) is dominant for the scattering process. 

The distributions of dusty gas around a massive star have not been understood well because of the complicated stellar feedback \citep[e.g.,][]{2007ARA&A..45..481Z, 2018MNRAS.473.4754F}. Therefore, we here investigate the impact of the configuration of dusty slabs on the CP map. 
In the phase distributions of $S_{41}/S_{11}$ for $a_{\rm d} = 1.0~{\rm \mu m}$, the maximum value appears at 
$90^{\circ} \lesssim \theta \lesssim 135^{\circ} $, corresponding to the back-scattering.
It means that the CP degree increases if dust grains are located behind the radiation source.
Figure \ref{zu4} shows the CP degrees in the cases that the dusty slab is shifted from the star to forward or backward by 0.1 pc.
As expected from Figure \ref{zu2}, the CP degree becomes larger if the slab is behind the radiation source.
For that case, the maximum of the CP degree reaches 60 percent.
On the other hand, it is lower if the slab is in front of the radiation source, and it becomes $\sim 30$ percent.
In the forward scattering, the component of $S_{41}/S_{11}$ becomes smaller, because there are photons with different signs of the CP.
 Therefore, we realize that the CP degrees change by a factor of $2-3$ depending on the configuration between a star and a dusty slab. 
Recent theoretical works suggest that the stellar feedback works anisotropically \citep[e.g.,][]{2017MNRAS.471.4844G, 2018ApJ...859...68K}. 
Hence, if the dusty gas along a line of sight is attenuated due to the feedback, we observe mostly back-scattered photons, which produce higher CP as shown in our simulations.

\section{Conclusions and Discussion}

We have explored the generation of CP at the wavelength of $2.14~{\rm \mu m}$ in star-forming regions using three-dimensional radiative transfer simulations. 
Assuming a dusty gas slab composed of aligned oblate grains, we have successfully reproduced the quadrupole patterns of CP, which is similar to the observed CP maps \citep{2010OLEB...40..335F}. 
We have found that the distributions and degrees of CP sensitively depend on the size of dust grains. 
The CP degree shows $|V|/I \sim 1$ percent for a grain size of $a_{\rm d} = 0.1~\rm \mu m$
and $|V|/I \sim 40$ percent for $a_{\rm d} = 1.0~\rm \mu m$. 
This can explain the observed variety of CP maps \citep{2014ApJ...795L..16K}.  
In previous works, the dichroic extinction model was suggested to produce the observed high degree of CP \citep{2013ApJ...765L...6K}. In that model, 
multiple scattering processes on several dust screens/clouds are requisite. 
In this {\it Letter}, we have demonstrated that a single scattering can produce the high degree of CP,
if micron-size grains are dominant in a dusty gas wall.

Recent observations show that the CP positively correlates with the luminosity of the central YSOs \citep{2014ApJ...795L..16K}.
High-mass YSOs are likely to form in massive high-pressure cores in molecular clouds \citep[e.g.,][]{2003ApJ...585..850M}, where the growth rate of dust size can be enhanced \citep{1978ppim.book.....S, 2013MNRAS.432..637A}. 
Therefore, we suggest that the observed correlation may be linked to the grain size distributions around the massive YSOs, although 
the growth rate of dust grains is still under the debate quantitatively.

In this study, we have also found that the CP degree is subject to the configuration between a star and a dusty slab. 
If a dusty slab is located behind the star, the CP degree becomes higher by a factor of $2-3$
compared to a slab in front of a star. 
Although we assumed a simple slab model in this study, the actual configuration is related to the evolution of the star-forming clouds. The directions of the minor axis of dust grains depend on the streaming of gas, magnetic, and radiation field \citep{2015ARA&A..53..501A}. Thus, to model the CP more precisely, we need to conduct magneto-radiation hydrodynamics simulations in star-forming clouds. 

In this work, we have considered the contribution of the silicate dust grains alone, but there are graphite ones in the interstellar medium.
The dielectric functions at the infrared wavelength are different between silicate and graphite grains \citep{2003ApJ...598.1017D}, and thus the CP degree can change with the composition of dust grains.
Besides, the bare silicate may not grow up to the micron size solely by coagulation \citep{2013MNRAS.434L..70H}.
The micron-size dust grains must contain buffers such as water ice.
We will investigate the dependence of the CP degrees on the composition of dust grains in the future works.

\section*{Acknowledgements}
The authors would like to thank Ryo Tazaki for fruitful discussions.
This research has been supported by Multidisciplinary Cooperative
Research Program at the Center for Computational Sciences,
University of Tsukuba,  MEXT as Exploratory Challenge on Post-K
Computer (Elucidation of the Birth of Exoplanets [Second Earth] and
the Environmental Variations of Planets in the Solar System), hp190185,
and partially by Grants-in-Aid for Scientific Research (JP19H00697, JP17H04827 and JP18H04570) 
from the Japan Society for the Promotion of Science (JSPS).




\bibliographystyle{mnras}

\input{paper1.bbl}


\bsp	
\label{lastpage}
\end{document}